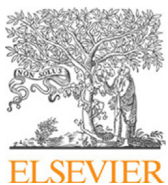
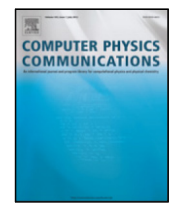

# Lab::Measurement—A portable and extensible framework for controlling lab equipment and conducting measurements[☆]

S. Reinhardt [a], C. Butschkow [a], S. Geissler [a], A. Dirnaichner [a,1], F. Olbrich [a,2], C.E. Lane [b], D. Schröer [c], A.K. Hüttel [a,*]

[a] *Institute for Experimental and Applied Physics, University of Regensburg, 93040 Regensburg, Germany*
[b] *Department of Physics, Drexel University, 3141 Chestnut Street, Philadelphia, PA 19104, USA*
[c] *Center for NanoScience and Department für Physik, Ludwig-Maximilians-Universität, Geschwister-Scholl-Platz 1, 80539 München, Germany*

**ABSTRACT**

Lab::Measurement is a framework for test and measurement automatization using Perl 5. While primarily developed with applications in mesoscopic physics in mind, it is widely adaptable. Internally, a layer model is implemented. Communication protocols such as IEEE 488 [1], USB Test & Measurement [2], or, e.g., VXI-11 [3] are addressed by the connection layer. The wide range of supported connection backends enables unique cross-platform portability. At the instrument layer, objects correspond to equipment connected to the measurement PC (e.g., voltage sources, magnet power supplies, multimeters, etc.). The high-level sweep layer automates the creation of measurement loops, with simultaneous plotting and data logging. An extensive unit testing framework is used to verify functionality even without connected equipment. Lab::Measurement is distributed as free and open source software.

**Program summary**
*Program Title:* Lab::Measurement 3.660
*Program Files doi:* http://dx.doi.org/10.17632/d8rgrdc7tz.1
*Program Homepage:* https://www.labmeasurement.de
*Licensing provisions:* GNU GPL v2[3]
*Programming language:* Perl 5
*Nature of problem:* Flexible, lightweight, and operating system independent control of laboratory equipment connected by diverse means such as IEEE 488 [1], USB [2], or VXI-11 [3]. This includes running measurements with nested measurement loops where a data plot is continuously updated, as well as background processes for logging and control.
*Solution method:* Object-oriented layer model based on Moose [4], abstracting the hardware access as well as the command sets of the addressed instruments. A high-level interface allows simple creation of measurement loops, live plotting via GnuPlot [5], and data logging into customizable folder structures.

[1] F. M. Hess, D. Penkler, et al., LinuxGPIB. Support package for GPIB (IEEE 488) hardware, containing kernel driver modules and a C user-space library with language bindings. http://linux-gpib.sourceforge.net/
[2] USB Implementers Forum, Inc., Universal Serial Bus Test and Measurement Class Specification (USBTMC), revision 1.0 (2003). http://www.usb.org/developers/docs/devclass_docs/
[3] VXIbus Consortium, VMEbus Extensions for Instrumentation VXIbus TCP/IP Instrument Protocol Specification VXI-11 (1995). http://www.vxibus.org/files/VXI_Specs/VXI-11.zip
[4] Moose—A postmodern object system for Perl 5. http://moose.iinteractive.com
[5] E. A. Merritt, et al., Gnuplot. An Interactive Plotting Program. http://www.gnuplot.info/

---

# 1. Introduction

Experimental physics frequently relies on complex instrumentation. While high-level equipment often has built-in support for automation, and while more and more ready-made solutions for complex workflows exist, the nature of experimental work lies in the development of new workflows and the implementation of new ideas. A freely programmable measurement control system is the obvious solution.

One of the most widely used applications of this type is National Instruments LabVIEW [1], where programming essentially means drawing the flowchart of your application. While this is highly flexible and supported by many hardware vendors, more complex use cases can easily lead to flowcharts that are difficult to understand and to maintain. A second widely used commercial application implementing instrument control is MathWorks MATLAB [2], where this ties into its numerical data evaluation and graphical user interface (GUI) capabilities.

`Lab::Measurement` provides a lightweight, open source alternative based on Perl 5, a well-known and well-established script programming language with a strong emphasis on text manipulation. `Lab::Measurement` has been successfully used for quantum transport measurements on GaAs/AlGaAs gate-defined quantum dots, see, e.g., [3–5], as well as carbon nanotube devices [6–8] and other mesoscopic systems [9,10].

As other open source measurement packages, see e.g. [11,12], `Lab::Measurement` features a modular structure which facilitates adding new hardware drivers. Here we highlight the unique features of `Lab::Measurement`. This includes in particular its high-level, descriptive "sweep" interface, where data and metadata preservation as well as real-time plotting is intrinsically provided. Notable is further the automated creation of unit tests via connection logging, and the use of traits [13] to exploit the modular structure of SCPI [14] and thereby minimize code duplication.

# 2. Implementation and architecture

`Lab::Measurement` is implemented as an object-oriented layer structure of Perl modules, as displayed schematically in Fig. 1. The figure does not show the full set of provided modules, but demonstrates a small selection of hardware.

The uppermost layer depicted in the figure, corresponding to the *hardware driver*, is not part of the `Lab::Measurement` Perl module distribution. Several Perl binding modules such as `Lab::VISA` (comparable to the Python PyVISA module [15]), `Lab::VXI11`, `Lab::Zhinst`, and `USB::TMC` are separately provided by the `Lab::Measurement` authors via the Comprehensive Perl Archive Network (CPAN), as bindings to third-party libraries. Conversely, e.g., the LinuxGPIB package [16] provides a Perl interface to its library and Linux kernel modules on its own.

The *connection layer* provides a unified API to these hardware drivers, making it possible to address devices, send commands, and receive responses. It handles access to many common automation protocols, such as GPIB, USB-TMC, VXI-11, raw TCP Sockets, or for example the Zurich Instruments LabOne API.

An overview of the currently supported connection types is given in Table 1.

The *instrument layer* represents connected test and measurement devices. The generic instrument class `Lab::Moose::Instrument` only provides means to send commands and receive responses; derived classes encapsulate the model-specific command sets. On this level, direct device programming can take place.

---

the Perl ecosystem. This means that it can be used and distributed according to the terms of either the GNU General Public License (version 1 or any later version) or the Artistic License; the choice of license is up to the user.

```perl
package Lab::Moose::Instrument::SR830;
use Moose;
use Lab::Moose::Instrument qw/validated_getter
                              validated_setter/;
use Lab::Moose::Instrument::Cache;

extends 'Lab::Moose::Instrument';
with qw(
  Lab::Moose::Instrument::Common
);
# the initialization
sub BUILD {
  my $self = shift;
  $self->clear(); $self->cls();
}
# reference signal frequency: declare cache
cache frq => ( getter => 'get_frq' );
# reference signal: get frequency
sub get_frq {
  my ( $self, %args ) = validated_getter( \@_ );
  return $self->cached_frq(
    $self->query( command => 'FREQ?', %args ) );
}
# reference signal: set frequency
sub set_frq {
  my ( $self, $value, %args )= validated_setter(
    \@_ , value => { isa => 'Num' } );
  $self->write(
          command => "FREQ $value", %args );
  $self->cached_frq($value);
}
# [...]
# measured value, get r and phi in one call
sub get_rphi {
  my ( $self, %args ) = validated_getter( \@_ );
  my $retval = $self->query(
              command => "SNAP?3,4", %args );
  my ( $r, $phi ) = split( ',', $retval );
  chomp $phi;
  return { r => $r, phi => $phi };
}
# [...]
```

Listing 1: Commented excerpt from the driver module for the Stanford Research SR830 lock-in amplifier; shown are the header and initialization, the functions for getting and setting the reference voltage frequency, and a function for reading out radius and phase of the measured signal.

The *sweep layer* finally provides abstraction of sweeping instrument parameters and performing multi-dimensional measurements, with live plotting and data logging. This facilitates the creation of complex measurement tasks for non-programmers without creation of loops or conditional control flow statements.

# 3. Extending and programming the instrument layer

## 3.1. Minimal example of an instrument driver

Listing 1 shows a commented excerpt of an instrument driver module, based on the Stanford Research SR830 lock-in amplifier module distributed with `Lab::Measurement`.

Low level functionality in drivers is frequently implemented via Moose roles. These are an implementation of the concept of traits, which enable horizontal code reuse more efficiently than single inheritance, multiple inheritance, or mixins [13,17]. In our case, roles are used to share functions between different instrument drivers. This significantly eases implementation of new drivers compared to other open-source measurement frameworks, e.g., [11,12]. The minimal driver of Listing 1 only consumes a single role, `Lab::Moose::Instrument::Common`, providing, e.g., the basic IEEE 488.2 standard commands. A more complex example demonstrating the usage of roles follows in Section 3.2.

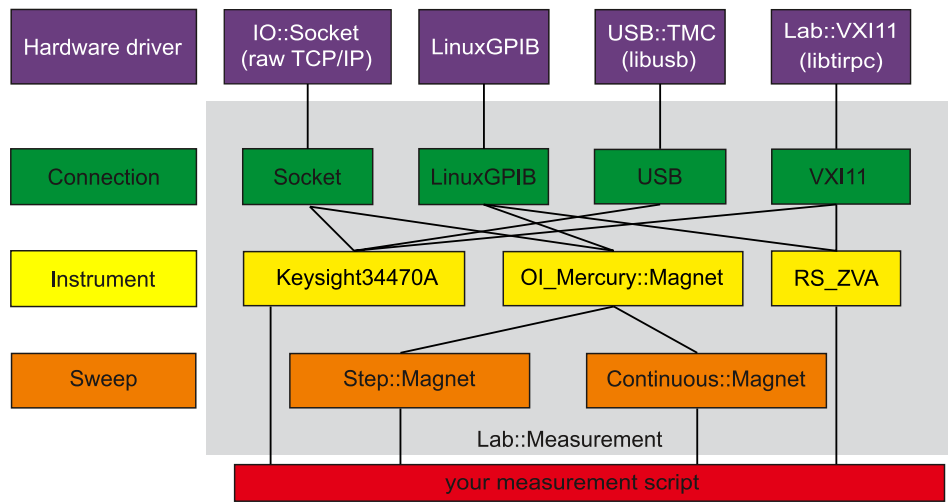

**Fig. 1.** Schematic of the internal structure of `Lab::Measurement`. The gray frame indicates modules that are part of the `Lab::Measurement` Perl distribution. For this example, as instrument hardware a Keysight 34470A digital multimeter, an Oxford Instruments Mercury iPS superconducting magnet controller, and a Rohde & Schwarz ZVA vector network analyzer are chosen.

**Table 1**
Available protocols and corresponding connection types. National Instruments (NI) VISA is only supported by NI on few Linux distributions and faces severe functional restrictions there due to the requirement of loading proprietary Linux kernel modules. In particular, USB GPIB adaptors are not supported by NI VISA on Linux.

| Protocol | Connection name | Operating system support | | Description |
|---|---|---|---|---|
| | | Linux | MS Windows | |
| Multiple | VISA | (X) | X | National Instruments VISA architecture device access |
| GPIB | LinuxGPIB | X | – | IEEE 488 / GPIB adaptors supported by the LinuxGPIB project |
| | VISA::GPIB | (X) | X | IEEE 488 / GPIB frontend for VISA |
| IP, TCP | Socket | X | X | IP network sockets, mostly used for TCP connections |
| VXI-11 | VXI11 | X | – | VXI-11 protocol via SunRPC or libtirpc |
| | VISA::VXI11 | X | X | VXI-11 frontend for VISA |
| USB T&M | USB | X | X | USB Test & Measurement class device driver using libusb |
| | VISA::USB | (X) | X | USB Test & Measurement frontend for VISA |
| ZI LabOne | Zhinst | X | X | Zurich Instruments LabOne API |

The code in the BUILD method is executed after instrument construction. Here we clear the device input and output buffers and any status or error flags. Otherwise we take care that driver initialization does not change any further device state; this is to ensure that, e.g., output voltages do not fluctuate on script start, or that manually applied configuration settings are kept unless explicitly overridden in the script.

The following code handles reading and setting the frequency of the ac voltage generated at the lock-in reference output. Hardware access is often slow, in particular regarding query operations, which is why we declare a cache for the frequency using functionality from `Lab::Moose::Instrument::Cache`. The cache keeps track of frequency values written to the instrument, for fast later access. This assumes that device access is exclusive to the running script and no modifications on, e.g., a hardware front panel are taking place.

`get_frq` is provided to obtain the ac frequency of the reference signal generated by the lock-in amplifier. It can be called with a hash of options as argument, which is validated by `validated_getter` and passed to the instrument query function. The queried value of the frequency is cached and returned to the caller. `set_frq` sets the reference output ac frequency of the lock-in amplifier. It is called with a hash, which must contain a key named `value` that provides the frequency value. The frequency is set by writing the corresponding command to the instrument; subsequently the cache is updated with the frequency value.

The last stored cache value of the frequency can be accessed by the user with the `cached_frq` method. If this method is called when the cache entry is still empty, the getter method declared with the `cache` keyword, `get_frq`, is called to populate the cache. In effect, `cached_frq` should be used in high-level functions whenever possible, except for situations where a hardware query needs to be enforced (refreshing the cache) with `get_frq`.

As opposed to the reference frequency, the signal amplitude and phase are not device parameters but measured values and thus cannot be set; the `get_rphi` function performs no caching. It returns a reference to a hash with keys `r` and `phi`. In general, a design choice of `Lab::Measurement` is that functions always return a scalar variable; either directly a scalar value as, e.g., a frequency or a measured current, or a reference to a more complex Perl object.

The full SR830 driver contains several additional but similarly implemented functions. This includes controlling reference ac voltage amplitude, time constant, filter settings, input configuration, and sensitivity as well as reading out the signal as $x$- and $y$-quadrature or even as simultaneous combination $(x, y, r, \varphi)$. The functions are documented in POD syntax; the documentation can, e.g., be read locally using perldoc or online on CPAN.

```perl
package Lab::Moose::Instrument::RS_FSV;
use 5.010;
use PDL::Core qw/pdl cat/;
use Moose;
use Moose::Util::TypeConstraints;
use MooseX::Params::Validate;
use Lab::Moose::Instrument qw/timeout_param
                              precision_param/;
use Carp;
use namespace::autoclean;

extends 'Lab::Moose::Instrument';
# ... and consumes the roles ...
with qw(
   Lab::Moose::Instrument::Common
   Lab::Moose::Instrument::SCPI::Format
   Lab::Moose::Instrument::SCPI::Sense::Bandwidth
   Lab::Moose::Instrument::SCPI::Sense::Frequency
   Lab::Moose::Instrument::SCPI::Sense::Sweep
   Lab::Moose::Instrument::SCPI::Initiate
   Lab::Moose::Instrument::SCPIBlock
);
# initialization
sub BUILD {
   my $self = shift;
   $self->clear();   $self->cls();
}
# get a spectrum trace
sub get_spectrum {
   my ($self, %args) = validated_hash(
       \@_,
       timeout_param(),
       precision_param(),
       trace => {isa => 'Int', default => 1},
   );
   # the FSV can store up to 6 traces
   my $trace = delete $args{trace};
   if ($trace < 1 || $trace > 6) {
     croak "trace has to be in (1..6)"; }
   # get frequency list (from SCPI::Sense::Freq)
   my $freq_array =
       $self->sense_frequency_linear_array();
   # make only one sweep (from SCPI::Initiate)
   if ( $self->cached_initiate_continuous() ) {
     $self->initiate_continuous( value => 0 ); }
   # single or double precision? (from SCPIBlock)
   my $precision = delete $args{precision};
   $self->set_data_format_precision(
     precision => $precision );
   # how much data will we read?
   my $num_points = @{$freq_array};
   $args{read_length} = $self->block_length(
       precision  => $precision,
       num_points => $num_points );
   # run sweep (from SCPI::Initiate)
   $self->initiate_immediate();
   # wait for completion (from Common, IEEE488.2)
   $self->wai();
   # request binary trace data
   my $binary = $self->binary_query(
       command => "TRAC? TRACE$trace",
       %args
   );
   # convert to PDL object
   my $points_ref = pdl $self->block_to_array(
       binary    => $binary,
       precision => $precision
   );
   # return it with frequency column prepended
   return cat((pdl $freq_array), $points_ref);
}
__PACKAGE__->meta()->make_immutable();
1;
```

Listing 2: Commented Rohde & Schwarz FSV series spectrum analyzer driver module.

### 3.2. Example instrument driver based on SCPI roles

Listing 2 provides an example of a more complex instrument driver module, at the example of a Rohde & Schwarz (R&S) FSV series spectrum analyzer. The driver strongly relies on the role functionality of Moose. The roles included in the R&S FSV driver with the `with qw(...);` statement mostly model the SCPI subsystems [14] which comprise the command syntax of the instrument.

The high-level `get_spectrum` method, central part of the driver module, initiates a frequency sweep, reads the resulting spectrum from the instrument, and returns the data as a 2D array to the user. In the following, we discuss at its example how the different code parts interact in this method.

The FSV spectrum analyzer can be configured to record and store up to 6 frequency sweeps (called traces); the `trace` parameter can be used to address one of these. The `precision` parameter selects between double and single floating point precision as defined in the SCPI standard. The variable `$freq_array` is initialized with an array of the frequency values addressed within a sweep; then the device is configured to perform only one sweep (instead of continuous, repeated measurement).

After calculating the size of the resulting data block, the sweep is started by calling the `initiate_immediate` method, provided by the `Lab::Moose::Instrument::SCPI::Initiate` role. Calling the `wai` method from `Instrument::Common`[4] (corresponding to the IEEE 488.2 command of the same name) waits until the operation has finished. Using `binary_query` from the base `Instrument` class, the measured data is read out once the sweep has completed.

The `block_to_array` method provided by the `Instrument::SCPIBlock` role is used to convert the raw IEEE 488.2 binary block data into a Perl array. The array is then subsequently converted into a PDL (Perl Data Language) object, combined with the frequency array, and returned to the caller of the `get_spectrum` method. The returned PDL object can then be used for further processing or can be written to a datafile.

### 3.3. Instrument layer usage

The instrument layer is the lowermost layer typically used in measurement scripts. All functions of an instrument driver can here directly be accessed, allowing fine-grained control and specialized programming. At the same time, the connection abstraction is present, allowing instrument control independent of the precise operating system of the control PC running Lab::Measurement or the wiring.

Listing 3 shows a brief example how to read out a Stanford Research SR830 lock-in amplifier connected via LinuxGPIB, using the driver discussed above. It sets the amplitude and frequency of the lock-in reference output. The new values are queried and printed together with the amplitude and phase of the measured signal.

Note that, as a consequence of abovementioned coding convention, `get_rphi` returns a scalar, namely a hash *reference*; the returned values are thus accessed as, e.g., `$rphi->{r}` and `$rphi->{phi}` .

## 4. Sweep layer usage

A complete example of a measurement script based on the Lab::Moose::Sweep interface is given in Listing 4 . It is intended for a magnetic resonance experiment, where, e.g., a ferromagnetic sample is mounted on a coplanar waveguide and microwave

---
[4] For convenience we subsequently shorten role names.

```perl
#!/usr/bin/env perl
# Read out SR830 lock-in amplifier
use 5.010;
use Lab::Moose;

my $lia = instrument(
    type             => 'SR830',
    connection_type  => 'LinuxGPIB',
    connection_options => {pad => 13}
);

$lia->set_amplitude(value => 0.5);
$lia->set_frq(value => 1000);

my $amp = $lia->get_amplitude();
say "Reference output amplitude: $amp V";

my $frq = $lia->get_frq();
say "Reference frequency: $frq Hz";

my $rphi = $lia->get_rphi();
say "Signal:  amplitude   r=$rphi->{r} V";
say "         phase     phi=$rphi->{phi} degree";
```

Listing 3: Accessing a Stanford Research SR830 lock-in amplifier connected via LinuxGPIB on the instrument class level: reference signal amplitude and frequency as well as signal amplitude and phase are read out and displayed.

transmission is measured as a function of applied magnetic field. The magnetic field is controlled with an Oxford Instruments (OI) Mercury iPS magnet controller; a Rohde & Schwarz (R&S) ZVA vector network analyzer (VNA) is used for generating the microwave signal and detecting the complex transmission parameter $S_{21}(f)$.

Part A of the listing defines and initializes the connected remote-controlled instruments in the same way as already demonstrated in Listing 3. Here, both instruments are connected via ethernet at fixed IP addresses, with the OI Mercury accepting plain TCP connections, and the R&S ZVA using VXI-11 protocol.

In part B, the two instruments are connected to so-called sweeps. The magnet controller is connected to a continuous magnet sweep, which is internally defined by a class `Lab::Moose::Sweep::Continuous::Magnet`. Multiple parameters such as start/stop points and the sweep rate are provided to the sweep function. The VNA is connected to a stepwise sweep of the output frequency. This way, the complex transmission will be measured at few discrete microwave frequencies; the VNA is doing fast "point measurements", instead of internally performing a frequency sweep.

Part C of Listing 4 defines the data file to be written. Data is recorded in the native GnuPlot [18] format as ASCII lines of tab-separated numerical values, with blank line separators after each conclusion of the innermost loop. The data file in addition contains a comment section with information on its columns; comment lines in the data file start with the hash character # and are ignored during plotting and evaluation.

In addition, here also a set of plot parameters of the data ("a plot") is defined, referencing the column names of the data file. The plot will be displayed during the measurement and refreshed after each completion of the innermost measurement loop. This functionality for foreground operation is purely optional, and implemented using GnuPlot calls. Future work shall address the creation of more complex graphical user interface components, see e.g. [11].

Part D defines the measurement subroutine, a callback that is executed for each point in the addressed output parameter space. In this example it reads out the momentary value of the magnetic field and performs a VNA "point measurement". As already noted, here the VNA is configured to only measure at a single frequency; other types of VNA measurements are also available. The returned $pdl object contains the frequency, real and imaginary parts of the $S_{21}$ transmission parameter, as well as amplitude and phase of

```perl
#!/usr/bin/env perl
use 5.010;
use Lab::Moose;
#--- A: Initialize the instruments -----------
my $ips = instrument(
    type             => 'OI_Mercury::Magnet',
    connection_type  => 'Socket',
    connection_options => {
        host => '192.168.3.15'
    },
);

my $vna = instrument(
    type             => 'RS_ZVA',
    connection_type  => 'VXI11',
    connection_options => {
        host => '192.168.3.27'
    },
);

# Set VNA's IF filter bandwidth (Hz)
$vna->sense_bandwidth_resolution(value => 1);
#--- B: Define the sweeps --------------------
my $field_sweep = sweep(
    type       => 'Continuous::Magnet',
    instrument => $ips,
    from       => 2,       # Tesla
    to         => 0,       # Tesla
    rate       => 0.01,    # Tesla/min
    start_rate => 1,       # Tesla/min
                           # (rate to approach
                           # start point)
    interval   => 0,       # run slave sweep
                           # as often as possible
);
# Measure complex transmission at 1,2,...,10GHz
my $frq_sweep = sweep(
    type       => 'Step::Frequency',
    instrument => $vna,
    from       => 1e9,
    to         => 10e9,
    step       => 1e9
);
#--- C: Create a data file and live plot -----
my $datafile = sweep_datafile(
    columns => [qw/B f Re Im r phi/]
);
# Add live plot of transmission amplitude
$datafile->add_plot(x => 'B', y => 'r');
#--- D: Measurement instructions -------------
my $meas = sub {
    my $sweep = shift;

    say "frequency f: ", $sweep->get_value();
    my $field = $ips->get_field();
    my $pdl = $vna->sparam_sweep(timeout => 10);
    $sweep->log_block(
        prefix => {field => $field},
        block  => $pdl
    );
};
#--- E: Connect everything and go ------------
$field_sweep->start(
    slave       => $frq_sweep,
    datafile    => $datafile,
    measurement => $meas,
    folder      => 'Magnetic_Resonance',
);
```

Listing 4: Example measurement script based on the sweep layer. An Oxford Instruments Mercury power supply is performing a continuous magnetic field sweep; in an inner loop, complex signal transmission is measured by a Rohde & Schwarz ZVA vector network analyzer at several discrete frequency values. Data is recorded in Gnuplot format and plotted in real time.

the $S_{21}$ parameter. This data is then logged with the `log_block` command, inserting the momentary magnetic field as additional column in front. The `start` method in part E finally starts the

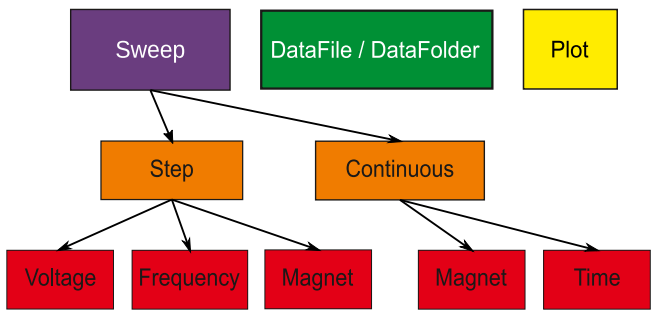

**Fig. 2.** Structural overview of the sweep and data logging classes.

```
---
id: 0
method: Clear
command: '*CLS'
id: 1
method: Write
command: '*RST'
id: 2
method: Write
timeout: 10
command: SNAP?1,2
id: 3
method: Query
retval: "0,0\n"
timeout: 3
```

Listing 5: Example YAML log file for mock connection testing of the SR830 lock-in amplifier driver.

measurement via the outer (magnet) sweep and provides the glue connecting sweeps, data file, and measurement subroutine.

The overall design of the `Lab::Measurement` code is intended to provide a flexible high-level interface, while requiring only a minimum of programming knowledge for typical measurement tasks. Given their structure, the measurement scripts might even be pregenerated by a graphical user interface, while still allowing a maximum of internal tuning and adaption for the expert. Fig. 2 shows an overview of the classes is the sweep layer and the data logging framework. Data files and live plots are implemented as separate classes, which can also be used independently of the sweep functionality.

## 5. Quality control

`Lab::Measurement` is provided with a set of unit tests that can be run upon installation. The unit tests are integrated with the Github development repository via Travis [19] (for testing on Linux, under a wide range of Perl 5 versions) and Appveyor [20] (on Microsoft Windows) and run immediately on addition of new code commits. The current test status of the master branch is displayed on the `Lab::Measurement` homepage. In addition, release test reports on a wide range of software combinations, submitted by volunteers, are available via the CPANtesters website [21].

### 5.1. Testing with mock connection objects

Comprehensive testing of a large part of the provided functionality is impossible without having the corresponding hardware connected. To extend testing as far as possible, our testing routines simulate instrument responses using mock connection objects as described below.

A mock test consists of an ordinary measurement script which interfaces with one or more instruments. To select between logging and mock operation, additional command line processing code is provided in the `MockTest` package, which is provided with the tests. The test script is first run connected to a real instrument. Logging of the communication is provided by the instrument layer `read`, `write`, and `query` methods, independent of the specific instrument driver module. Each call to these methods is recorded with its arguments and the resulting return values into a human readable YAML text file, which is then stored alongside with the test script. An example log file for the SR830 lock-in amplifier is shown in Listing 5 .

Running a test without the connected instrument will use the `Mock` connection, which takes such a YAML log file as input. Method calls on the mock connection will be compared with the recorded method calls; deviations from the log will lead to a test failure, allowing to ascertain the impact of committed code on the device communication. In case the test script provides the correct commands, the respective instrument responses are provided from the log file. This framework allows the creation of unit tests with considerably more coverage compared to those distributed in alternative open-source solutions [11,12].

## 6. Historical interface

Over the past years, `Lab::Measurement` has undergone several large scale revisions. Development was started in 2006, initially only as a solution for addressing devices via the National Instruments (NI) VISA libraries [22,23] for Linux or MS Windows. The code was later extended and refactored into a layer model, providing the abstractions for connections to instruments and instrument model specific features. The initial high-level interface focused only on logging and plotting, without providing any loop functionalities. Its successor, `Lab::XPRESS`, was added in 2012 and is with its documentation still present in the distribution for ongoing support of measurement scripts. The intent is to remove this deprecated code when all parts have been fully ported to Moose and `Lab::Measurement` 4 is released.

The historical stack uses an additional bus layer, in analogy to a GPIB adaptor carrying connections to several instruments. All components are implemented using minimal object-oriented Perl. In the high-level interface, `Lab::XPRESS`, sweep classes, data logging and plotting are tightly integrated. A main objective of the Moose-based sweep classes was to modularize and simplify this part of the software, in order to allow easier integration of new features.

Aspects that still require porting to the new high-end Moose code include predefined command line switch handling for measurement scripts and customized handling of keyboard interrupts.

## 7. Requirements and licensing

`Lab::Measurement` is written in pure Perl and designed to run on any version since Perl 5.14. Given the extensive use of Moose and its object-oriented programming features, a range of Perl module dependencies is required. These, including minimum versions where applicable, are declared in the distribution metadata `META.json` file included in the package and can be installed automatically by `cpan` or any Perl package manager.

The package is tested regularly and supported on both Linux and Microsoft Windows, however, given availability of hardware back-ends, should also work on other operating systems. Hardware back-ends may require further dependencies, as, e.g., in the case of `Lab::VISA` the National Instruments VISA driver stack, or in the case of `USB::TMC` the `libusb` system library. Further details on these requirements can be found in the documentation of the back-end modules on CPAN.

Given that any plotting or graphical interface is optional and that the CPU and memory footprint of running Perl scripts is very low compared to commercial instrument control solutions [1,2], `Lab::Measurement` also lends itself for background control operation, logging, and headless server operation. The open source nature of its dependencies particularly on Linux allows the use arbitrary Linux systems, in particular including, e.g., ARM or AArch64 based embedded architectures.

As many other Perl module distributions, `Lab::Measurement` is released under the same license as Perl itself, meaning dual-licensed either under the Artistic License or the GNU General Public License, version 1 or any later version published by the Free Software Foundation. With this it is free and open source software, and can be downloaded from the Comprehensive Perl Archive Network (CPAN).

## 8. Summary

`Lab::Measurement` is a measurement framework designed for high portability and extensibility. The framework is in active use and was applied successfully in many condensed matter physics experiments. Its implementation consists of an object oriented layer structure. The connection layer provides backends for common instrumentation protocols on both Linux and Microsoft Windows. The instrument layer, which represents connected equipment, uses modern object oriented programming techniques for code reuse. A high-level interface, suitable for users without previous Perl experience, allows fast creation of complex measurement tasks involving continuous or discrete sweeps of control parameters as, e.g., voltages, temperature, or magnetic field. By using mock connection objects, unit tests can be run without connected instruments. This allows safe refactoring and facilitates the future evolution of the framework.

## Acknowledgments

We would like to thank E. E. Mikhailov for his helpful comments and considerable recent code contributions, and K. Fredric for insightful discussions. The authors thank the Deutsche Forschungsgemeinschaft for financial support via SFB 689, GRK 1570, and Emmy Noether grant Hu 1808/1.